\documentclass[floats,twocolumn]{revtex4}

\usepackage{graphicx}
\usepackage{amsfonts}

\newcommand{\ZZ}{\mathbb{Z}}

\newcommand{\DeclareMathOperator}[2]{\def#1{\mathop{\mathrm{#2}}\nolimits}}

\newcommand{\cccc}[4]{\ensuremath{\hat{c}_{#1} \hat{c}_{#2} \hat{c}_{#3} \hat{c}_{#4}}}

\newcommand{\ccp}[2]{\ensuremath{\hat{c}_{#1} \hat{c}'_{#2}}}

\def\be{\begin{equation}}
\def\ee{\end{equation}}
\def\ba#1{\begin{array}{#1}}
\def\ea{\end{array}}
\def\bn{\begin{enumerate}}
\def\en{\end{enumerate}}

\def\H{\mathcal{H}}

\def\beq{\begin{equation}}
\def\eeq{\end{equation}}

\def\Wt{W_{\text{tot}}}
\def\Vt{V_{\text{tot}}}

\begin{document}

\title{The effects of interactions on the topological classification of free fermion systems}
\author{Lukasz Fidkowski}
\affiliation{California Institute of Technology, Pasadena, CA 91125, U.S.A.}
\author{Alexei Kitaev}
\affiliation{California Institute of Technology, Pasadena, CA 91125, U.S.A.}

%\pacs{}
%\date{\today}
\begin{abstract}
We describe in detail a counterexample to the topological classification of free fermion systems.  We deal with a one dimensional chain of Majorana fermions with an unusual $T$ symmetry.  The topological invariant for the free fermion classification lies in $\mathbb{Z}$, but with the introduction of interactions the $\mathbb{Z}$ is broken to $\mathbb{Z}_8$.  We illustrate this in the microscopic model of the Majorana chain by constructing an explicit path between two distinct phases whose topological invariants are equal modulo $8$, along which the system remains gapped.  The path goes through a strongly interacting region.  We also find the field theory interpretation of this phenomenon.  There is a second order phase transition between the two phases in the free theory which can be avoided by going through the strongly interacting region.  We show that this transition is in the $2D$ Ising universality class, where a first order phase transition line, terminating at a second order transition, can be avoided by going through the analogue of a high temperature paramagnetic phase.  In fact, we construct the full phase diagram of the system as a function of the thermal operator (i.e. the mass term that tunes between the two phases in the free theory) and two quartic operators, obtaining a first order Peierls transition region, a second order transition region, and a region with no transition.
\end{abstract}

\maketitle

\section{Introduction \label{intro}}

The discovery of the quantum spin Hall effect \cite{KaneMele0,KaneMele1,HgTe0,HgTe1} and of the strong 3D topological insulator \cite{FuKaneMele,Roy,FuKane-BiSb,BiSb}, both of which are novel band insulators, has prompted renewed interest in the study of topological phases of free fermion systems.  Indeed, a full classification of all possible topological phases in such systems has been put forward in \cite{SRFL}, where it is related to the enumeration of symmetry classes of matrices \cite{AZ}, and in \cite{kitaev-2009}, which uses the mathematical machinery of $K$-theory.  This classification is rather successful, with physical representatives of the non-trivial topological classes listed for dimensions $1$, $2$, and $3$, including the quantum spin Hall system HgTe and the 3D topological insulator BiSb.

The big open question now is how the presence of interactions changes this classification.  Specifically, it is possible that phases that were distinct in the free classification can actually be adiabatically connected through a strongly interacting region. Now, for certain systems the topological invariants can be defined in terms of physically measurable quantities, and hence are stable to interactions.  This occurs for example in the integer quantum Hall effect, where the integer Chern number is proportional to the Hall conductivity, as well as in 2D chiral superconductors and 2D topological insulators and superconductors.  Also, the $\ZZ_2$ classification of the $3D$ topological insulator reflects the presence or absence of a $\pi$ theta term in the effective action for the electromagnetic field, extending the definition of this invariant to include systems with interactions \cite{classInsulators,essin-2008}.

In this paper we give an example where the free classification breaks down.  The system is $1$ dimensional, with an unusual $T$ symmetry: $\hat{T}^2=1$ instead of
$\hat{T}^{2}=(-1)^{\hat{N}}$.   For a concrete model, we consider the Majorana
chain and its variations, where $\hat{T}$ acts on odd sites by
$\hat{T}\hat{c}_{j}\hat{T}^{-1}=-\hat{c}_{j}$ so that terms like
$i\hat{c}_{j}\hat{c}_{k}$ are only allowed between sites of different
parity. In the free-fermion setting, this symmetry is described by one
positive Clifford generator, hence $p=-1$,\,\, $q=p+2=1$, and for $d=1$ we get
a topological invariant $k\in\pi_{0}(R_{q-d})=\ZZ$.

We can get some intuition for this integer by thinking about boundary states.   We start by comparing it to the usual $\ZZ_2$ classification of 1D systems without symmetry.  The $\ZZ_2$ classification of systems without symmetry is reflected in the fact that for a pair of Majorana chains $\hat{c}_j^\alpha$, $\alpha = 1,2$, we can gap out the dangling Majorana operators $\hat{c}_1^\alpha$ and $\hat{c}_N^\alpha$ at the ends of the chain by introducing the terms $i \hat{c}_1^1 \hat{c}_1^2$ and $i \hat{c}_N^1 \hat{c}_N^2$ (the $i$ is necessary to make these terms Hermitian).  However, these terms are not $\hat{T}$-invariant: $\hat{T} (i \hat{c}_j^1 \hat{c}_j^2) \hat{T}^{-1} = -i \hat{c}_j^1 \hat{c}_j^2$.  Thus, in the $\hat{T}$-symmetric case, we cannot gap out the dangling Majorana operators with quadratic interactions, for any number of chains - this is the origin of the $\ZZ$ invariant, which just counts the number of boundary states in this setup.  However, it turns out that we can use non-trivial quartic interactions to gap out the dangling Majorana modes for the case of $8$ Majorana chains - this is what we will focus on in this paper.

Thus we study the setting of $8$ parallel Majorana chains, which has a phase transition characterized by $k=8$.  We will see how the two phases separated by this transition are actually adiabatically connected through an interacting phase.  This means that the two phases are actually the same, and that the $\ZZ$ topological invariant is actually broken down to $\ZZ_8$.  Below we will demonstrate this fact by constructing an explicit path in Hamiltonian space connecting the two phases of $8$ parallel Majorana chains.  Adiabatic transformation along this path connects the two phases through a strongly interacting, but everywhere gapped, region.  We do the analysis first for the microscopic model in section \ref{mm}, where we construct a quartic $\hat{T}$-invariant interaction that gaps out the $8$ boundary Majoranas, and then for the continuum theory in section \ref{ca}.  

\section{The microscopic model}\label{mm}

We consider $8$ parallel Majorana chains $\hat{c}^{\alpha}_j$, where $\alpha=1,\ldots,8$ is the chain index.  The $T$ symmetry still acts by $\hat{T}\hat{c}^{\alpha}_{j}\hat{T}^{-1}=-\hat{c}^{\alpha}_{j}$.  The Hamiltonian is the sum of the Hamiltonians for the individual chains:

\begin{equation}
\hat{H}=\sum_{\alpha=1}^8 \hat{H}_{\alpha}
\end{equation}

\begin{equation}
\hat{H}_{\alpha}=\frac{i}{2}
\left(u\sum_{l=1}^{n}\hat{c}^{\alpha}_{2l-1}\hat{c}^{\alpha}_{2l}
+v\sum_{l=1}^{n-1}\hat{c}^{\alpha}_{2l}\hat{c}^{\alpha}_{2l+1}\right).
\end{equation}We construct a path from a representative Hamiltonian of the $u<v$ phase to one of the $u>v$ phase.  
While in principle we could start with any representative Hamiltonians for the two phases, it will be especially convenient to choose so-called fully dimerized ones, i.e. $u=1,v=0$ and $u=0,v=1$.  This choice turns off the odd (or even) bond couplings and thus breaks down the chains into easy to analyze independent finite size systems.  For example, for $u=1,v=0$, the finite dimensional sub-systems consist of the Majoranas $\{ \hat{c}^{1}_{2l-1}, \hat{c}^{1}_{2l}, \ldots, \hat{c}^{8}_{2l-1}, \hat{c}^{8}_{2l} \}$, and for $u=0,v=1$ they consist of $\{ \hat{c}^{1}_{2l}, \hat{c}^{1}_{2l+1}, \ldots, \hat{c}^{8}_{2l}, \hat{c}^{8}_{2l+1} \}$.  Both are $256$ dimensional and we will generically denote their Hilbert space ${\cal H}_0$.

The key idea is now to work with these finite dimensional systems, in order to make a fully analytic treatment possible.  Indeed, to connect the two phases, we start with one fully dimerized Hamiltonian, say $u=1,v=0$, and turn on an interaction $W$ which contains quartic terms, but only ones that are products of Majoranas with the same site index (but different chain indices), i.e. terms of the form $c^{\alpha_1}_l c^{\alpha_2}_l c^{\alpha_3}_l c^{\alpha_4}_l$.  The virtue of such an interaction is precisely that it does not couple the finite dimensional sub-systems; also, as required, it is $T$-invariant.  We show that as we ramp up our specially constructed $W$, we can turn off the kinetic terms entirely (i.e. turn off $u$ so that both $u,v=0$) while maintaining a gap in ${\cal H}_0$ (and hence in the full system).  We then reverse the procedure, turning on the opposite kinetic term $v$, and turning off $W$ to get the opposite dimerized Hamiltonian $u=0,v=1$.

\subsection{Construction of the interaction term W}

We have described the path in Hamiltonian space that will connect the two phases; all that is left now is to construct the interaction term $W$ in such a way that as $W$ is turned on and the kinetic term turned off, the gap in ${\cal H}_0$ is maintained.  We have thus reduced the problem to a completely finite dimensional one.

$W$ must couple different chains but only at the same site; thus it is made up of $8$ Majorana fermions $\hat{c}_1, \ldots, \hat{c}_8$.  To motivate the construction, we must first delve a bit into some representation theory.  In general, we can consider $2n$ Majorana's $\hat{c}_1,\ldots,\hat{c}_{2n}$ forming a $2^n$ dimensional Hilbert space ${\cal H}$ (so in our case of interest $n=4$).  ${\cal H}$ is a representation of $so(2n)$ as follows: given a skew-symmetric matrix $A \in so(2n)$, we let

\begin{equation}
\rho (A)=\frac{i}{4} \sum_{j,k=1}^{2n} A_{jk} \hat{c}_j \hat{c}_k
\end{equation}The $i$ in front is to make the matrix Hermitian, in order to obtain a unitary representation.  It is easy to check that $\left[ -i \rho (A), -i \rho (B) \right] = -i \rho ([A,B])$, so that $\rho$ defines a map of Lie algebras.   The induced action on the $\hat{c}_i$,

\begin{equation}
\hat{c}_l \rightarrow i [ \rho(A), c_l]
\end{equation} is just the standard $\hat{c}'_l = \sum_j A_{lj} c_j$.  

Before considering $so(8)$, we warm up by studying $so(4)$.  Note that $so(4)=so(3) \oplus so(3)$.  Under $\rho$, the generators of the two $so(3)$'s are

\begin{equation}
\left(
\frac{i}{2} (\hat{c}_1 \hat{c}_2 - \hat{c}_3 \hat{c}_4),\,
\frac{i}{2} (\hat{c}_1 \hat{c}_4 - \hat{c}_2 \hat{c}_3),\,
\frac{i}{2} (\hat{c}_1 \hat{c}_3 + \hat{c}_2 \hat{c}_4) \right)
\end{equation} and

\begin{equation}
\left(
\frac{i}{2} (\hat{c}_1 \hat{c}_2 + \hat{c}_3 \hat{c}_4),\,
\frac{i}{2} (\hat{c}_1 \hat{c}_4 + \hat{c}_2 \hat{c}_3),\,
\frac{i}{2} (-\hat{c}_1 \hat{c}_3 + \hat{c}_2 \hat{c}_4) \right)
\end{equation}The condition for a state $|\psi\rangle \in {\cal H}$ to be annihilated by all generators of the first $so(3)$ is equivalent to $\hat{c}_1 \hat{c}_2 \hat{c}_3 \hat{c}_4 |\psi\rangle = - |\psi\rangle$, and there are two states that satisfy this condition, forming a spin-$1/2$ representation of the other $so(3)$.  Similarly, $\hat{c}_1 \hat{c}_2 \hat{c}_3 \hat{c}_4 |\psi\rangle = |\psi\rangle$ for $|\psi\rangle$ annihilated by the second $so(3)$.  Indeed, under $so(3) \oplus so(3)$, ${\cal H}$ decomposes as:

\begin{equation}
{\cal H} = (0,\frac{1}{2}) \oplus (\frac{1}{2}, 0)
\end{equation}

From this analysis, we can already see that the above strategy for constructing the path through the space of gapped Hamiltonians connecting the two phases would fail for the case of $4$ Majorana chains.  This is because for $4$ Majorana chains the only possible quartic interaction $W$ is proportional to $\hat{c}_1 \hat{c}_2 \hat{c}_3 \hat{c}_4$, and regardless of its sign leaves a doubly degenerate ground state for each group of $4$ corresponding sites, leading to a gapless system at the midpoint of the path when the kinetic terms are turned off.  Indeed, a field theory analysis shows that the corresponding marginal quartic interaction in the continuum limit is a sum of two quartic terms, one for each $so(3)$, but with opposite signs.  One of these is irrelevant and the other relevant, leading to one of the $so(3)$ sectors being gapped out but the other remaining, producing the (gapless) antiferromagnetic Heisenberg model.  We will have more to say about the field theory analysis in the next section.

The analysis of $4$ chains does however lead to a natural way to gap out the $8$ chain system.  Indeed, we can split the $8$ chains up into $2$ groups of $4$, turn on a quartic interaction in each group of $4$ to create low energy spin $1/2$'s as in the above paragraph, and then couple these spin $1/2$'s into a non-degenerate singlet via an anti-ferromagnetic interaction.  We will now flesh out this intuition and give an explicit construction of $W$.

To facilitate the analysis, we make use of the $so(8)$ symmetry.  Per the construction above, we have $8$ Majoranas forming a $16$ dimensional representation $so(8)$.  This representation is actually a direct sum of two spinor representations, $8_{+}$ and $8_{-}$, of $so(8)$, both of which are equivalent to the $8$ vector representation in the sense that there is a group of so-called triality automorphisms of $so(8)$ \cite{Shankar} which interchange the $8_{+}$, $8_{-}$, and $8$.  Now, while it turns out that a $W$ that is fully $so(8)$ symmetric doesn't work, we can find one that is symmetric under the $so(7) \subset so(8)$ that fixes a particular element of say $8_{+}$.  This is the triality conjugate of one of the familiar $so(7)$ subalgebras.

The goal for the remainder of this sub-section is to show that the following explicit expression for $W$
\begin{eqnarray}\label{W}
W &=&\cccc{1}{2}{3}{4} + \cccc{5}{6}{7}{8} +\cccc{1}{2}{5}{6} \nonumber \\
     &+&\cccc{3}{4}{7}{8} - \cccc{2}{3}{6}{7} -\cccc{1}{4}{5}{8} \nonumber \\
     &+&\cccc{1}{3}{5}{7} + \cccc{3}{4}{5}{6} +\cccc{1}{2}{7}{8} \nonumber \\
     &-&\cccc{2}{3}{5}{8} - \cccc{1}{4}{6}{7} +\cccc{2}{4}{6}{8} \nonumber \\
     &-&\cccc{1}{3}{6}{8} - \cccc{2}{4}{5}{7}
\end{eqnarray} is $so(7)$ invariant.  The proof that turning on $W$ and turning off the kinetic terms leaves ${\cal H}$ gapped will be left for the next sub-section.

%It is useful to have a geometric interpretation of these terms.  Figure WHATEVER displays the $8$ Majoranas at the vertices of a cube.  The quartic terms in $W$ then correspond to all possible coplanar sets of the $4$ vertices (note that here we are dealing with a vector space over $\mathbb{Z}_2$).

To show that $W$ is $so(7)$-invariant, we first have to specify the $so(7)$ we are talking about.  To do this, we first combine the Majoranas into regular fermions as follows:

\begin{eqnarray}
\hat{c}_{2j-1} &=& (a_j + a^{\dag}_j)  \\
\hat{c}_{2j} &=& -i (a_j - a^{\dag}_j) 
\end{eqnarray}We thus have $4$ regular fermions.  Let $|0\rangle$ be the state where all of them have occupation number $0$, and let $|\psi\rangle = \frac{1}{\sqrt{2}} (|0\rangle - a^{\dag}_1 a^{\dag}_2 a^{\dag}_3 a^{\dag}_4 |0\rangle)$.  $|\psi\rangle$ is a spinor in $8_{+}$, and we claim that it is the unique ground state of $W$, and that $W$ is invariant under the $so(7)$ subgroup of $so(8)$ that leaves $|\psi\rangle$ fixed.

Let us start by proving $|\psi\rangle$ is the unique ground state of $W$.  To do this, we first consider

\begin{equation}
W_1 = \cccc{1}{2}{3}{4}+\cccc{5}{6}{7}{8}+\cccc{1}{2}{5}{6}+\cccc{1}{3}{5}{7}
\end{equation} Note that each of the $4$ terms in $W_1$ has eigenvalues $\pm 1$.  Re-writing the terms in $W_1$ in terms of creation and annihilation operators, it is easy to see that $|\psi\rangle$ is an eigenstate of eigenvalue $-1$ for all of them.  It is also easy to see explicitly that $|\psi\rangle$ is the only state with this property - assuming eigenvalue $(-1)$ for the first three terms immediately gives a state that is a linear combination of $|0\rangle$ and $a^{\dag}_1 a^{\dag}_2 a^{\dag}_3 a^{\dag}_4 |0\rangle$, and the correct linear combination yielding $|\psi\rangle$ is fixed by the last term.

The key point now is that each term in $W$ can be written as a product of terms in $W_1$ - this immediately shows that $|\psi\rangle$ is an eigenstate of each term of $W$, with eigenvalue $\pm 1$.  In fact, it can be explicitly checked that all the eigenvalues are $-1$, so that $|\psi\rangle$ is a ground state of $W$.  Uniqueness follows from the fact that it is already a unique ground state of $W_1$.

Now that we have proved that $|\psi\rangle$ is the unique ground state of $W$, we can finally show that $W$ is invariant under the $so(7)$ that leaves $|\psi\rangle$ fixed.  First, we identify the generators of $so(7)$.  Note that there are $28$ linearly independent bilinears $\hat{c}_i \hat{c}_j$.  Take any one of the $14$ terms in $W$, say $\cccc{i}{j}{k}{l}$, transposing a pair of $\hat{c}$'s if necessary to make the sign positive.  We then claim that the set of all bilinears $\frac{i}{2} (\hat{c}_i \hat{c}_j -\hat{c}_k \hat{c}_l), \frac{i}{2} (\hat{c}_i \hat{c}_l -\hat{c}_j \hat{c}_k),\frac{i}{2} (\hat{c}_i \hat{c}_k +\hat{c}_j \hat{c}_l)$, as $\cccc{i}{j}{k}{l}$ ranges over the $14$ terms in $W$, spans $so(7)$.  To see this, first notice that all these bilinears annihilate $|\psi\rangle$ - this follows from the fact that $|\psi\rangle$ is an eigenvector of eigenvalue $-1$ for the corresponding terms in $W$.  Thus the bilinears span a subset of $so(7)$.  To see that they actually span all of $so(7)$ we use a dimension argument - by brute force we compute the rank of the relevant $28$ by $28$ matrix and find it equal to $21$, the dimension of $so(7)$.  With this description of $so(7)$, checking the invariance of $W$ amounts to computing its commutator with all the generators of $so(7)$.  This is actually rather tractable, because for any given generator constructed in this paragraph, its commutator with all but $4$ terms of $W$ is trivially $0$, and can be easily done via computer.

\subsection{Adiabatic continuation using W}

We now explicitly verify that we can connect the two phases adiabatically.  As explained, this amounts to showing that we can turn on a quartic interaction and turn off the kinetic terms completely while maintaining a gap in the $256$ dimensional Hilbert space ${\cal H}_0$.  That is, we have the $16$ Majorana fermions $\hat{c}_i, \hat{c}'_i$, $i=1,\ldots, 8$, and consider the following interactions:

\begin{eqnarray}
\Wt &=& W+W' \\
T &=&  \sum_{i=1}^8 i \ccp{i}{i} \\
H &=& w \, \Wt + t \, T
\end{eqnarray}where $W$ is as above, and $W'$ is given by replacing all the $\hat{c}_i$'s with $\hat{c}'_i$'s.  Thus $T$ is the kinetic term and $W$ the quartic potential.

%We show that the Hamiltonian $H = t T + (1-t) \Wt$ stays gapped for all $0 \leq t \leq 1$.

\begin{figure}[htp]
\includegraphics[width=7.5cm]{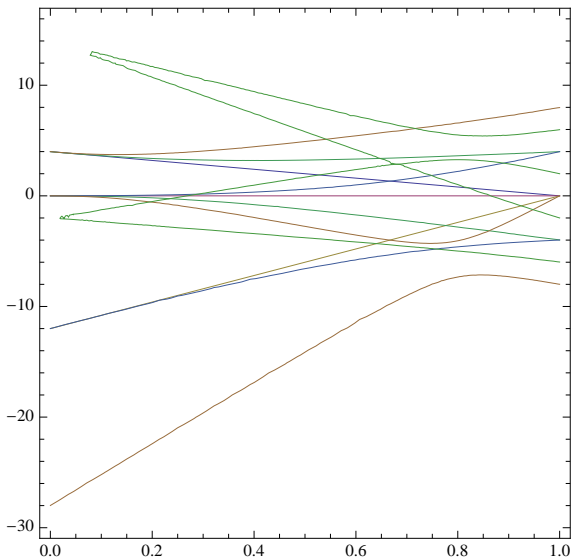}
\caption{Eigenvalues of $H = t \, T + (1-t) \, \Wt$ as a function of $t$.  The system remains gapped throughout the path. \label{eplot}}
\end{figure}

We have explicitly diagonalized this Hamiltonian numerically.  The eigenvalues are plotted in fig. \ref{eplot}, where we take the path $w=1-t, t \in [0,1]$.  As we can see, the system remains gapped throughout.

However, it is nice to also have a clear analytical argument that the system
remains gapped.  To do this, let us first analyze the symmetries.  The $256$
dimensional Hilbert space ${\cal{H}}_0$ for these $16$ Majorana fermions has
an action of the Lie algebra $so(8) \oplus so(8)$, with one $so(8)$ acting on
the $\hat{c}_i$'s and the other on the $\hat{c}'_i$'s.  The potential term
$\Wt$ is invariant under an $so(7) \oplus so(7)$ subalgebra, whereas the
kinetic term $T$ is invariant under the diagonal $so(8)$.  In fact, to
construct an easy to analyze Hamiltonian, it will be useful to add in a fully
$so(8) \oplus so(8)$ invariant quartic term $\Vt = V+V'$.  $V$ and $V'$ are
quartic in the fermions like $W$ and $W'$, and are proportional to the
quadratic Casimirs of the corresponding $so(8)$'s (plus some constant
term); we will set the constant of proportionality below.  The Hamiltonian $\tilde{H}$ thus is

\begin{equation}\label{Hamil}
\tilde{H} = t \, T + w \, \Wt + v\, \Vt
\end{equation} The plan for the remainder of this section is to use the symmetries to reduce the problem of finding the spectrum of $\tilde{H}$ to something more manageable.  We first study the actions of $T$, $\Wt$, and $\Vt$ separately, and then block decompose $\tilde{H}$, with the blocks corresponding to different representations of the diagonal $so(7)$, which is a symmetry of all $3$ terms, and hence $\tilde{H}$.  Then we construct a path through $(t, w, v)$ space which connects $(1,0,0)$ and $(-1,0,0)$.

A convenient basis for understanding better the action of $T$ is as follows.  We define $b_i, b_i^\dag$ through

\begin{eqnarray}{\label{des1}}
\hat{c}_i &=& (b_i + b_i^\dag) \\
\hat{c}_i' &=& -i (b_i - b_i^\dag)
\end{eqnarray}We note that each term $i \ccp{i}{i} = b_i^\dag b_i +b_i b_i^\dag$ is equal to $\pm 1$, depending on the occupation number of the $b$ fermion on site $i$.  Thus the possible eigenvalues of $T$ are $\pm 8, \pm 6, \pm 4, \pm 2$ and $0$.  Under the diagonal $so(8)$, the corresponding eigenspaces have dimension $8$-choose-$k$ where $k$ is the number of occupied sites, i.e.: $1, 8, 28, 56$ and $70$.  It will be convenient to consider superpositions of the $m$ and $-m$ eigenvalues of $T$ and deal with representations $1_L, 8_L, 28_L, 56_L$ and $1_R, 8_R, 28_R, 56_R$, where $L$ and $R$ are exchanged under the action of $T$.  All of these are irreducible, whereas $70 = 35^s \oplus 35^v$, where $35^s$ and $35^v$ are distinct $35$ dimensional representations.

Now that we understand the action of $T$, let us figure out the actions of $\Wt$ and $\Vt$.  As a representation of $so(8) \oplus so(8)$, we can think of ${\cal{H}}_0$ as a tensor product of two $16$ dimensional Hilbert spaces, one for the $\hat{c}_i$'s, and one for the $\hat{c}'_i$'s (in fact, it is a graded tensor product, but this distinction will not make a difference in our analysis).  This decomposition is useful because it is preserved by $\Wt$.  As representations of $so(8)$ we also have $16 = 8_{-} \oplus 8_{+}$ where $8_{-}$ and $8_{+}$ are the two different chirality spinors.  These are distinguished by the sign of the fermion parity operator, with $8_{+}$ having even fermionic parity, and $8_{-}$ having odd parity.  Indeed, we have two fermion parity operators, $(-1)^{F_L} = \prod_{i=1}^8 \hat{c}_i$ and $(-1)^{F_R} = \prod_{i=1}^8 \hat{c}'_i$, as well as the total fermion parity operator $(-1)^F = (-1)^{F_L + F_R}$.  The potential terms $\Wt$ and $\Vt$ respect both parities, whereas the kinetic term respects only $(-1)^F$, and flips $(-1)^{F_L}$ and $(-1)^{F_R}$.  

We can now expand the tensor product:

\begin{eqnarray} {\label{des2}}
&&{\cal H}_0 = (8_{-} \oplus 8_{+}) \otimes (8_{-} \oplus 8_{+}) = \\
&&=(8_{-} \otimes 8_{-}) \oplus (8_{-} \otimes 8_{+}) \oplus (8_{+} \otimes 8_{-}) \oplus (8_{+} \otimes 8_{+}) \nonumber
\end{eqnarray} With this description of the Hilbert space, it is easy to
figure out the actions of $\Wt$ and $\Vt$.  Indeed, $\Wt$ is a sum of two
terms, $W$ and $W'$, that act independently on the left and right hand factors
in the tensor product.  Both $W$ and $W'$ are quadratic in the Lie algebra
generators and commute with $so(7)$ (which means they are proportional to the
quadratic Casimir of $so(7)$, plus a possible constant).  We note that under
the $so(7)$, $8_{-} \rightarrow 8$ and $8_{+} \rightarrow 1 \oplus 7$.  Using
these facts and doing some computation, we find that $W$ annihilates $8_{-}$,
has eigenvalue $-14$ on the state $|\psi\rangle \in 8_{+}$ that is fixed by
the $so(7)$, and has eigenvalue $2$ on the remaining vector multiplet $7$ of
so(7), and similarly for $W'$.  The actions of $V$ and $V'$, which are $so(8)$
invariant, are even simpler: they assign a different energy to $8_{+}$ and
$8_{-}$.  We choose the coefficients so that this energy is equal to $0$
for $8_{+}$ and $1$ for $8_{-}$.

We now want to relate the two descriptions (\ref{des1}) and (\ref{des2}) of ${\cal H}_0$ to find the action of $\tilde{H}= t\, T + w\, \Wt + v\, \Vt$.  As a first approximation to a connecting path, we attempt to connect the phases in a purely $so(8)$ invariant way.  That is, we connect the points $(1,0,0)$ and $(-1,0,0)$ in the $(t,w,v)$ space of Hamiltonians (\ref{Hamil}) by varying only $t$ and $v$.  Specifically, the path is \begin{equation}
\left(t,w,v \right) = \left(\cos \theta, 0, \sin\theta\right), \,\theta \in [0,\pi] \end{equation} The computation of the spectrum along this path reduces, by $so(8)$ symmetry, to diagonalizing the blocks corresponding to the $1, 8, 28, 56$, and $35$ representations.

Now, we have the following decomposition of the tensor factors in (\ref{des2}) as representations of the diagonal $so(8)$:

\begin{eqnarray}
8_{-} \otimes 8_{-} &=& 1 \oplus 28 \oplus 35^{ s } \label{line1} \\
8_{-} \otimes 8_{+} &=& 8 \oplus 56 \label{line2} \\
8_{+} \otimes 8_{-} &=& 8 \oplus 56 \label{line3} \\
8_{+} \otimes 8_{+} &=& 1 \oplus 28 \oplus 35^{ v } \label{line4}
\end{eqnarray}Let us figure out the action of $T$.  By looking at the fermion number, we see that $T$ exchanges (\ref{line1}) and (\ref{line4}), and (\ref{line2}) and (\ref{line3}).  The eigenvalues of $T$ on the pairs of $1,8,28,56$, and $35$ representations are $\pm 8, \pm 6, \pm 4, \pm 2$, and $0$, and $v \,V$ just assigns an energy $v$ to (\ref{line2}) and (\ref{line3}), and an energy $2v$ to (\ref{line1}).  So $\tilde{H}$ is represented by the following $2$ by $2$ matrices on these pairs of representations:

\begin{equation} \label{matrix1}
1: \left( \begin{array}{cc}
0 & 8 \,\cos \theta \\
8 \,\cos \theta & 2 \,\sin \theta \end{array} \right)
\end{equation}

\begin{equation} \label{matrix8}
8: \left( \begin{array}{cc}
\sin \theta & 6 \,\cos \theta \\
6 \,\cos \theta & \sin \theta \end{array} \right)
\end{equation}

\begin{equation} \label{matrix28}
28: \left( \begin{array}{cc}
0 & 4 \,\cos \theta \\
4 \,\cos \theta & 2 \,\sin \theta \end{array} \right)
\end{equation}

\begin{equation} \label{matrix56}
56: \left( \begin{array}{cc}
\sin \theta & 2 \,\cos \theta \\
2 \,\cos \theta & \sin \theta \end{array} \right)
\end{equation}with the two $35$'s splitting up into the $1$ by $1$ matrices $2\, \sin \theta$ and $0$ respectively.  Diagonalizing these, we see that the lowest eigenvalue occurs in the $1$ sector; it is $\sin \theta - \sqrt{8 \cos^2 \theta + \sin^2 \theta}$.

%We start by decomposing the tensor factors in (\ref{des2}) as representations of the diagonal $so(7)$, since $T$, $\Wt$, and $\Vt$ all commute with the diagonal $so(7)$ \cite{Slansky}:

%\begin{eqnarray}
%8_{-} \otimes 8_{-} &=& 1 \oplus (7 \oplus 21) \oplus 35 \label{line1} \\
%8_{-} \otimes 8_{+} &=& 8 \oplus (8 \oplus 48) \label{line2} \\
%8_{+} \otimes 8_{-} &=& 8 \oplus (8 \oplus 48) \label{line3} \\
%8_{+} \otimes 8_{+} &=& 1 \oplus (7 \oplus 21) \oplus (1 \oplus 7 \oplus 27) \label{line4}
%\end{eqnarray}Here we have bracketed terms that constitute diagonal $so(8)$ multiplets.

The problem with this path is that there is a large degeneracy at $\theta =
\pi / 2$, where the lowest eigenvalues for the $1, 28$, and the $35$
representations are $0$.  To break this degeneracy, we will turn on the
$so(7)$-symmetric $\Wt$ term.  We can work locally around $\theta = \pi/2$,
where (with a suitable rescaling) we can set
$v=\sin\theta\approx1$ ,
and keep $t=\cos\theta$ infinitesimal.  With such infinitesimal $t$, the
space of low energy states is well approximated for our purposes by $8_{+}
\otimes 8_{+} = 1 \oplus 28 \oplus 35$,  see (\ref{line4}). To second order
in $t$, the eigenvalues of $V + t\,T$ are $-32 t^2$ on the $1$, $-8t^2$ on the $28$, and $0$ on $35$.  To understand the action of $\Wt$ on $8_{+} \otimes 8_{+}$, we note

\begin{eqnarray}
8_{+} \otimes 8_{+} &=& (1 \oplus 7) \otimes (1 \oplus 7) \nonumber \\
&=& 1 \otimes 1 \, \oplus\, (1 \otimes 7 \oplus 7 \otimes 1) \,\oplus\, (7 \otimes 7).
\end{eqnarray}  From the discussion above, we know that the action of $\Wt$ on
this decomposition is as follows: $\Wt$ has eigenvalue $-28w$ on $1 \otimes
1$, $-12w$ on $1 \otimes 7 \oplus 7 \otimes 1$, and $4w$ on $7 \otimes 7$.  To
relate this decomposition to (\ref{line4}), we first use the fact that under
$so(7)$, $28 \rightarrow 7 \oplus 21$, and $35 \rightarrow 1 \oplus 7 \oplus
27$ \cite{Slansky}.  Now, $7 \otimes 7 = 1 \oplus 21 \oplus 27$.  Let
$\tilde{H} = V + t\, T + w \, \Wt$.  We immediately have the eigenvalues $4w$
on the $27$ and $4w - 8t^2$ on $21$. Also, the two $7$'s have the
eigenvalues $-12w$ and $-12w-8t^2$.  All that is left is to compute the
eigenvalues of $\tilde{H}$ on the two $1$'s, and show that one of them is
lower than the other and any of the above.

\begin{figure}[htp]
\includegraphics[width=7.5cm]{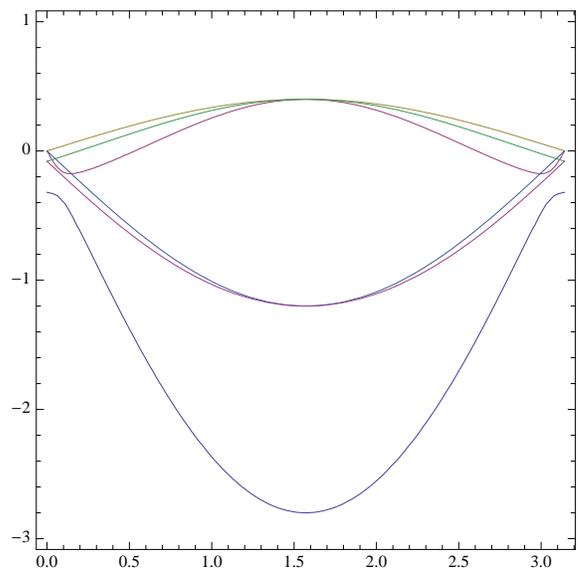}
\caption{Eigenvalues of $\tilde{H}=V + t\,T + w\,\Wt$ along the path $(t,w) = (0.1 \cos \theta, 0.1 \sin \theta), \, \theta \in [0, \pi]$.  The degeneracy is broken and the system remains gapped. \label{eplot17}}
\end{figure}

To do this, let us denote the $1$ in $8_{+} = 1 \oplus 7$ by $| \psi \rangle$,
and let $| \psi^{2} \rangle \equiv | \psi \rangle \otimes |\psi
\rangle$.  Then the $so(8)$ invariant $1$ in $8_{+}\oplus 8_{-}$ can be written as

\begin{equation}
|1_{so(8)}\rangle = \frac{1}{\sqrt{8}} \left( |\psi^{2} \rangle + \sum_{j=1}^7 |\xi_j \rangle \otimes |\xi_j \rangle \right)
\end{equation} Let $P \equiv |1_{so(8)}\rangle \langle 1_{so(8)} |$.  The effective Hamiltonian on the two dimensional space of $1$'s is then

\begin{equation}
H_{\text{eff}} = - 32 t^2 P  - 28 w |\psi^{2} \rangle \langle
\psi^{2} | + 4w\left(1-|\psi^{2}\rangle\langle\psi^{2}| \right), 
\end{equation} which is represented in matrix form by \begin{equation} \label {matrixH} \left( \begin{array}{cc} -28w -4t^2 & -4 \sqrt{7} t^2 \\ -4 \sqrt{7} t^2 & 4w-28t^2 \end{array} \right). \end{equation}  We graph the eigenvalues of this matrix, as well as those for the $7$, $21$, and $27$ representations, in figure \ref{eplot17}.

\section{Continuum analysis}\label{ca}

We now turn to a field theory analysis of the above phenomenon.  In the free system, the transition is simply that of $8$ parallel decoupled Majorana chains, so its field theory description is

\begin{eqnarray} \label{freeH}
 \hat{H}_{\text{free}} &\equiv&\hat{H}_0 + \hat{H}_m \\
&=&\frac{i}{2}\,\sum_{j=1}^8 \int(\eta_j \partial \eta_j - \bar{\eta}_j \partial \bar{\eta}_j) \,dx \\
            &+&\frac{i}{2} \,\sum_{j=1}^8 \int m\, \eta_j \bar{\eta}_j \, dx.
\end{eqnarray}Here the mass $m$ is a free parameter, and the phase transition occurs at $m=0$, where the system is described by the conformal field theory (CFT) of $8$ free Majorana fermions, or, equivalently, the $SO(8)_1$ WZW model.  We have shown in the previous section how to smoothly connect the phase with positive $m$ to one with negative $m$ through a strongly interacting region in a microscopic lattice model.  In this section, we construct a field theory analogue of that path.  

%The interaction $\Wt$ that we turned on in the previous section was quartic, so we consider quartic interactions in the fields $\eta_j$.

To study the field theory of the transition, we will need to understand the structure of the critical point in more depth.  As we said, this is just the CFT of $8$ free Majorana fermions, which, in addition to the fermions also contains spin fields, and these will be useful in our constructions.  For the case of one free fermion, there are the conformal weight $(1/16, 1/16)$ order and disorder operators $\sigma (z, \bar{z})$ and $\mu (z, \bar{z})$.  For the case of $8$ free fermions, we can form the $2^8 = 256$ possible products of $\sigma_i$ and $\mu_i$ ($i = 1, \ldots, 8$).  These have conformal weights $(1/2,1/2)$ and are precisely (linear combinations of) the $16^2$ primary fields $ \psi_i \bar{\psi}_j$, $\chi_i \bar{\psi}_j$, $\psi_i \bar{\chi}_j$, $\chi_i \bar{\chi}_j$, $i,j=1,\ldots, 8$, where $\psi$, $\chi$, $\bar{\psi}$, $\bar{\chi}$ are the two chiral left and right moving spinor representations of $SO(8)$ - indeed, this is a manifestation of the equivalence between the WZW $SO(8)_1$ and free fermion models \cite{CFT}.

The $SO(8)_1$ WZW model has a triality symmetry which permutes the $\eta, \psi$, and $\chi$ fields.  The fact that the Hamiltonian (\ref{freeH}) is invariant under triality, as well as other facts we will need, can be seen from bosonizing the system.  We pair up the $8$ Majorana fermions $\eta_j$ into $4$ Dirac fermions and bosonize those:

\begin{eqnarray} \label{bos}
\eta_{2j-1} \pm i \eta_{2j} &=& \exp \, ({\pm i \phi_j})\\
\bar{\eta}_{2j-1} \pm i \bar{\eta}_{2j} &=& \exp \, ({\pm i
\bar{\phi}_j})
\end{eqnarray}
The $16$ $(1/2, 1/2)$ fields

\begin{equation} \label{bosonizedspinors}
\exp {\frac{i}{2} (\pm \phi_1 \pm \phi_2 \pm \phi_3 \pm \phi_4)}
\end{equation}are then the spinors $\psi$ and $\chi$, with chirality distinguished by the parity of the number of minus signs in (\ref{bosonizedspinors}).  By doing a change of coordinates in $\vec \phi$ space, we can go to a basis in which the spinors look like the vectors, etc. - indeed, we can implement the entire triality symmetry group by appropriate rotations in $\vec \phi$ space.

Now that we have an understanding of the critical point, we look at perturbing it with the quartic interaction $W$ (\ref{W}) from the previous section.  Let us first determine the continuum limit of this interaction.  From the action $\hat{c}^i_j \rightarrow (-1)^j \hat{c}^i_j$ of the $T$ symmetry we see that at low energies

\begin{eqnarray}
\hat{c}^i_{2k} &=& \eta_i (2k) + {\bar{\eta}}_i (2k) \\
\hat{c}^i_{2k-1}&=&\eta_i (2k-1) - {\bar{\eta}}_i (2k-1)
\end{eqnarray} with $T$ just swapping $\eta_i$ and ${\bar{\eta}}_i$.  Now, $W$ is a sum of $14$ quartic terms; to start off we analyze just one of these, and we will take the sum over two adjacent sites: $\hat{c}^1_{2k-1} \hat{c}^2_{2k-1} \hat{c}^3_{2k-1} \hat{c}^4_{2k-1} + \hat{c}^1_{2k} \,\hat{c}^2_{2k} \,\hat{c}^3_{2k} \,\hat{c}^4_{2k}$.  We have

\begin{eqnarray}
\hat{c}^1_{2k-1} \hat{c}^2_{2k-1} \hat{c}^3_{2k-1} \hat{c}^4_{2k-1} + \hat{c}^1_{2k}\, \hat{c}^2_{2k}\, \hat{c}^3_{2k}\, \hat{c}^4_{2k} = \\
\prod_{i=1}^4 \left( \eta_i + {\bar{\eta}}_i \right) + \prod_{i=1}^4 \left( \eta_i - {\bar{\eta}}_i \right)
\end{eqnarray}Expanded out, this gives $8$ terms, $6$ of which are nonchiral, with the other $2$ being products of all $\eta$'s or all ${\bar{\eta}}$'s.  Instead of trying to argue that these last $2$ are irrelevant, we will simply ignore them - after all, we have the freedom to choose whatever continuum Hamiltonian we wish.  The remaining $6$ terms can be reorganized suggestively as follows:

\begin{eqnarray}
\frac{1}{2} \left( \eta_1 \eta_2 + \eta_3 \eta_4 \right) \left({\bar{\eta}}_1 {\bar{\eta}}_2 + {\bar{\eta}}_3 {\bar{\eta}}_4 \right) \\
-\frac{1}{2} \left( \eta_1 \eta_2 - \eta_3 \eta_4 \right) \left({\bar{\eta}}_1 {\bar{\eta}}_2 - {\bar{\eta}}_3 {\bar{\eta}}_4 \right) \\
+\frac{1}{2} \left( \eta_1 \eta_4 + \eta_2 \eta_3 \right) \left({\bar{\eta}}_1 {\bar{\eta}}_4 + {\bar{\eta}}_2 {\bar{\eta}}_3 \right) \\
-\frac{1}{2} \left( \eta_1 \eta_4 - \eta_2 \eta_3 \right) \left({\bar{\eta}}_1 {\bar{\eta}}_4 - {\bar{\eta}}_2 {\bar{\eta}}_3 \right) \\
+\frac{1}{2} \left( \eta_1 \eta_3 - \eta_2 \eta_4 \right) \left({\bar{\eta}}_1 {\bar{\eta}}_3 - {\bar{\eta}}_2 {\bar{\eta}}_4 \right) \\
-\frac{1}{2} \left( \eta_1 \eta_3 + \eta_2 \eta_4 \right) \left({\bar{\eta}}_1 {\bar{\eta}}_3 + {\bar{\eta}}_2 {\bar{\eta}}_4 \right)
\end{eqnarray}This can be compactly rewritten as
\begin{eqnarray}
 \frac{1}{2}\, \sum_{a=1}^3 J_a {\bar{J}}_a - \frac{1}{2} \, \sum_{a=1}^3 J'_a {\bar{J}}'_a
\end{eqnarray} with $J_a$ and $J'_a$ being the generators of the two $SU(2)$ groups in the $SO(4)$ of the $4$ chains.

Now, when we take the sum over all $14$ such quartic terms in $W$, we get a more complicated expression.  However, we know that it is a sum of terms of the form $\eta_i \eta_j {\bar{\eta}}_i {\bar{\eta}}_j$, i.e. a sum of products of right moving and left moving $so(8)$ currents.  We also know that it is $so(7)$ invariant, where the $so(7)$ leaves a {\it spinor} $\psi_8$ fixed.  In addition, we know that it is $T$ invariant.  Using the fact that the $so(8)$ currents can be equally well expressed as bilinears of the spinors $\psi_j$ (this follows from bosonization (\ref{bos}, \ref{bosonizedspinors})), we see then that the only allowed interactions are of the form

\begin{equation} \label{int}
H_{\text{int}}= -A \, \left( \sum_{j=1}^7 \psi_j \bar{\psi}_j \right)^2 - B \, \left( \sum_{j=1}^7 \psi_j \bar{\psi}_j \right) \psi_8 \bar{\psi}_8
\end{equation}This follows from the fact that, with the $T$ symmetry, the allowed interactions are quadratic forms on $so(8)$.  As a representation of $so(7)$, $so(8) = so(7) \oplus 7$, and both of these irreducible factors have exactly one invariant quadratic form.  

For the remainder of the analysis, it will be convenient to use triality to re-express the free Hamiltonian (\ref{freeH}) in terms of the $\psi$, i.e. treat the $\psi$ as the fundamental fields.  Without a mass term, it is:

\begin{equation}
\hat{H}_0=\frac{i}{2}\,\sum_{j=1}^8 \int(\psi_j \partial \psi_j - \bar{\psi}_j \partial \bar{\psi}_j) \,dx 
\end{equation}We now view both (\ref{int}) and the mass term $m \sum_{j=1}^8 \eta_j \bar{\eta}_j$ as interactions, and construct a $3$ dimensional phase diagram in $A$, $B$, and $m$ (the parameters $A$, $B$, and $m$ will have non-trivial renormalization group flows, and for the purposes of the phase diagram are all defined at some fixed energy scale).  To do this, we first need to express the mass term in terms of the free fermions $\psi_j$.  If we look at the $\psi_j$ as fundamental vectors, the $\eta_j$ and $\bar{\eta}_j$ are then spinors and can be expressed as linear combinations of products of order and disorder operators.  Comparing the bosonized form of the order and disorder operator bilinears \cite{IZ,Tsvelik}

\begin{eqnarray}
\sigma_{2k-1} \, \sigma_{2k} = \sin \frac{ \phi_k-\bar{\phi}_k }{2} \nonumber \\
\mu_{2k-1} \, \mu_{2k} = \cos \frac{ \phi_k-\bar{\phi}_k }{2}
\end{eqnarray}and the bosonization of the mass term via (\ref{bos},
\ref{bosonizedspinors}) we can show that (up to a constant factor)

\begin{equation} \label{smprod}
 \frac{i}{2} \sum_{j=1}^8 \eta_j \bar{\eta}_j = -\prod_{j=1}^8 \sigma_j + \prod_{j=1}^8 \mu_j
\end{equation}Note that here the $\sigma_j$ and $\mu_j$ are order and disorder operators for the $\psi_j$ fermions.

The total Hamiltonian is thus
\begin{eqnarray}
\hat{H}&=&\frac{i}{2}\,\sum_{j=1}^8 \int(\psi_j \partial \psi_j - \bar{\psi}_j \partial \bar{\psi}_j) \,dx \\
&-& \,  m  \int \left( \prod_{j=1}^8 \sigma_j - \prod_{j=1}^8 \mu_j \right) \, dx \\
&-&A \, \left( \sum_{j=1}^7 \psi_j \bar{\psi}_j \right)^2 - B \, \left( \sum_{j=1}^7 \psi_j \bar{\psi}_j \right) \psi_8 \bar{\psi}_8
\end{eqnarray} The goal for the remainder of this section will be to show that we can smoothly connect the region $m>0$ to the region $m<0$ by turning on $A$ and $B$.  To accomplish this, we first work out some features of the phase diagram - the full phase diagram is shown in figure (\ref{flow2}).  We can get some information about the phase diagram by studying specific special points.  Indeed, the line $B=2A$, ($A>0$) gives the marginally relevant interaction

\begin{equation}
-A \left( \sum_{j=1}^8 \psi_j \bar{\psi}_j \right)^2 = -A \left( \sum_{j=1}^8 \eta_j \bar{\eta}_j \right)^2.
\end{equation}This is the Gross-Neveu model, which is known to be integrable and gapped, and in which $\langle \eta_i \bar{\eta}_j\rangle = M \delta_{ij}$ condenses, spontaneously breaking the ${\mathbb{Z}}_2$ chiral symmetry $\eta_j \bar{\eta}_j \rightarrow - \eta_j \bar{\eta}_j$ (indeed, the spectrum of the model and the action of triality on the different types of excitations is well understood - see \cite{Shankar}).  The two vacua have opposite signs of $M$, and are adiabatically connected to the $+m$ and $-m$ gapped phases respectively.  Thus, to connect the $+m$ and $-m$ phases, all we have to do is connect these two vacua, which we can think of as corresponding to infinitesimal points $m=\pm \epsilon$.

The line $B=-2A$, ($A>0$) gives another Gross-Neveu point, related to the previous one by the chiral transformation $\psi_8 \rightarrow - \psi_8$, $\bar{\psi}_8 \rightarrow \bar{\psi}_8$, and the lines $B=2A$ and $B=-2A$ for $A<0$ give rise to marginally irrelevant interactions that flow to the gapless $SO(8)_1$ theory.  Also, the line $B=0$ gives rise to the $so(7)$ symmetric Gross-Neveu model

\begin{equation}
-A \left( \sum_{j=1}^7 \psi_j \bar{\psi}_j \right)^2
\end{equation}

This line provides the most important insight into the problem.  It has a gapless Majorana mode, $\psi_8$ and we can explore its neighborhood by turning on an infinitesimal $B$:

\begin{equation}
\H_{\text{int}} = -A \left( \sum_{j=1}^7 \psi_j \bar{\psi}_j \right)^2 - B \, \left( \sum_{j=1}^7 \psi_j \bar{\psi}_j \right) \psi_8 \bar{\psi}_8
\end{equation} with $B \ll A$.  We can solve this Hamiltonian by treating the
$B$ term as a perturbation.  Thus we have an expectation value $\sum_{j=1}^7
\psi_j \bar{\psi}_j = M'$, and the effective low energy theory is simply that
of the free fermion $\psi_8$ with mass $B M'$:

\begin{eqnarray} \label{Heff}
H_{\text{eff}} &=& \frac{i}{2}\, \int(\psi_8 \partial \psi_8 - \bar{\psi}_8 \partial \bar{\psi}_8) \,dx \\
&-&  m \left(\langle \sigma \rangle^7 \, \sigma_8 -\langle \mu \rangle^7 \,
\mu_8\right) - B M' \, \psi_8 {\bar{\psi}_8}
\end{eqnarray}
There are two possibilities: if the condensate $M'=\sum_{j=1}^7 \psi_j
\bar{\psi}_j >0$ then the $\sigma_j, j= 1,\ldots, 7$ gain expectation values
and $m$ couples to the Ising order parameter $\sigma_8$, whereas if $M'<0$,
then $\mu_j, j=1,\ldots,7$ gain expectation values and $\mu_8$ plays the role
of the order parameter. Either way, the low- and high-temperature phases of the
effective Ising model are realized at $B>0$ and $B<0$ respectively.

\begin{figure}[htp]
\includegraphics[width=8cm]{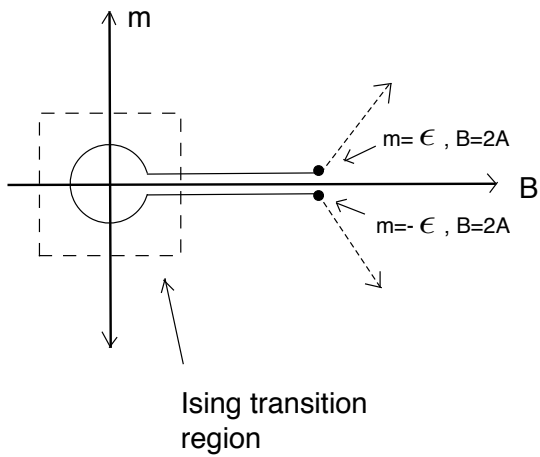}
\caption{Smoothly connecting the $m>0$ and $m<0$ phases through an Ising transition region}\label{fig2}
\end{figure}

We now have enough information to connect the two phases - see figure
(\ref{fig2}).  First of all, the two phases are connected to the $m=\pm
\epsilon$ points on the $B=2A, A>0$ line.  From the RG diagram (see discussion
below), unless something unexpected happens at strong coupling we see that
these points are adiabatically connected to the points $0<B \ll A$, $m=\pm
\epsilon$.  But here we can work with the effective Hamiltonian (\ref{Heff}).
If we go from $(B,+\epsilon)$ to $(B, -\epsilon)$ by going around the origin
in the $(B,m)$ plane, through the $B<0$ region (so as to avoid the positive
$B$ axis), the quantity $Z = \frac{i}{2}\sum_{j=1}^8 \eta_j \bar{\eta}_j =
-\prod_{j=1}^8 \sigma_j + \prod_{j=1}^8 \mu_j$ (see (\ref{smprod})) will vary
smoothly. This is the desired path that connects the two vacua.  Thus,
because $B$ controls the thermal operator and $m$ the order (or disorder)
field, the critical theory turns out to be in the universality class of the
$2D$ Ising transition.

For completeness we now work out the entire renormalization group flow diagram.  The two operators

\begin{eqnarray}
X &=& \left( \sum_{j=1}^7 \psi_j \bar{\psi}_j \right)^2 \\
Y &=& \left( \sum_{j=1}^7 \psi_j \bar{\psi}_j \right) \psi_8 \bar{\psi}_8
\end{eqnarray}form a closed set under RG flow, as they are the only $so(7)$ invariant $(1,1)$ operators available.  To track the flow of the coupling constants $A$ and $B$ we use Polyakov's formula \cite{Tsvelik}

\begin{equation}
\frac{d g_n}{d \ln (\Lambda/k)} = -2 \pi C^{pq}_n g_p g_q + {\cal O} (g^3)
\end{equation}where the $C^{pq}_n$ are the coefficients of the three point function of the corresponding operators.  Thus we are reduced to computing $3$ point functions of products of $X$ and $Y$.  This is trivial, since we are in a free fermion theory and can apply Wick's theorem.  We immediately see that, since $Y$ is the only operator containing $\psi_8$, we must have an even number of $Y$'s for a nonzero coefficient.  So the only nonzero coefficients are $C^{XX}_X, C^{XY}_Y, C^{YX}_Y,$ and $C^{YY}_X$.  The RG flow equations then read

\begin{eqnarray}
\frac{dA}{d \ln (\Lambda/k)} &=& S A^2 + T B^2 \\
\frac{dB}{d \ln (\Lambda/k)} &=& U AB
\end{eqnarray} for some constants $S, T, U>0$.  Because $B=2A$ is a fixed line, $U=S+4T$.  We do not calculate $S$ and $T$; all positive $S$ and $T$ give qualitatively similar results for the RG flow diagram, which is reproduced in figure (\ref{flow}) below.

\begin{figure}[htp]
\includegraphics[width=7.5cm]{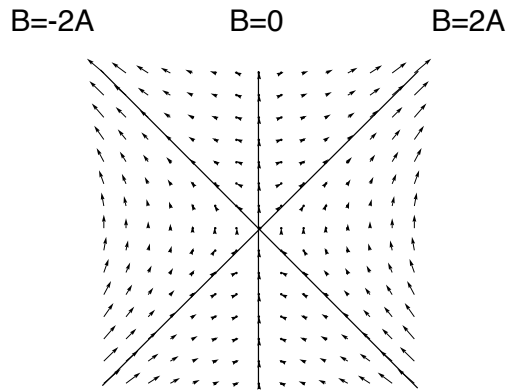}
\caption{RG flow for $B$ ($x$-axis) and $A$ ($y$-axis). \label{flow}}
\end{figure}

\begin{figure}[htp]
\includegraphics[width=7.5cm]{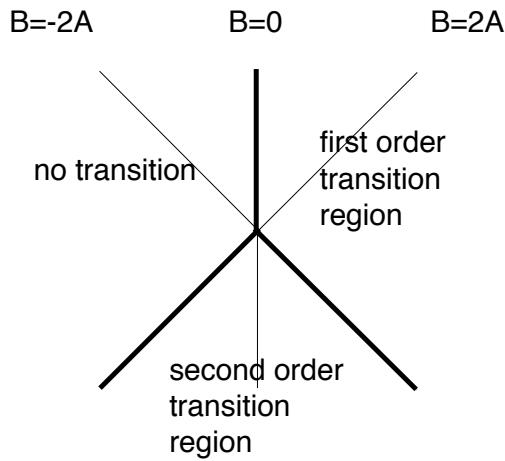}
\caption{Phase diagram for the transition from negative to positive $m$, indicating dependence on $B$ ($x$-axis) and $A$ ($y$-axis)
 \label{flow2}}
\end{figure}

From the flow diagram (\ref{fig2}), we can construct an entire phase diagram for the transition from positive to negative $m$.  Indeed, all points with $B>0$ that are above the line $B=-2A$ flow to the same fixed point and correspond to a first order transition.  As we saw, there is no transition in the $B<0$ region, and this extends to all points above the line $B=2A$.  Finally, the remaining points flow to the origin, which corresponds to free fermions and thus indicates a second order transition.  The situation is summarized in figure (\ref{flow2}).

\acknowledgments We would like to acknowledge useful discussions with John
Preskill and Andreas Ludwig.  This work was supported in part by the
Institute for Quantum Information under National Science Foundation grant
no. PHY-0803371.  A.~K. is also supported by DARPA grant
no. HR0011-09-0009.

\bibliographystyle{aipproc-atitle}

\vspace{5mm} 
\bibliography{mrefs1}

\end{document}